\documentclass[aps,prb,twocolumn,groupedaddress,showpacs,letterpaper]{revtex4}
\usepackage{amsmath}
\usepackage{graphicx}

\begin{document}

\title{Nanotube bundles and tube-tube orientation: A van der Waals Density Functional Study} 

\author{Heiko Dumlich\ref{aff:FUBerlin}}
\email[Corresponding author: ]{heiko.dumlich@fu-berlin.de}

\author{Stephanie Reich\ref{aff:FUBerlin}}

\affiliation{\label{aff:FUBerlin} Fachbereich Physik, Freie Universit\"{a}t Berlin,
14195 Berlin, Germany}

\date{\today}

\begin{abstract}
We study the binding energy, intertube distance and electronic structure of bundles consisting of 
single walled carbon nanotubes of the same chirality. We model various nanotube structures 
(chiralities) and orientations with van der Waals density functional theory. The orientation of 
the tubes in the bundle strongly influences the properties of the bundles if the chirality of the 
tubes shares symmetry with the trigonal bundle structure, meaning chiralities which have a 
$60^\circ$-rotational symmetry (C$_6$-axis), e.g. ($12,0$) bundles. The bundle structure breaks 
the symmetry depending on the arrangement of the neighboring tubes. Pseudogaps open in the 
electronic density of states and intertube distances ($\pm5$-$10\%$) vary in dependence of the 
relative orientation of the tubes in the bundle. Bundles of C$_6$-axis armchair tubes have 
metallic configurations. A $15^\circ$ rotation off the high symmetry configuration (AA-stacked) 
of a ($6,6$) bundle shows metallic behavior and has higher binding energy than the high symmetry 
configuration. We find binding energies between $19~\mbox{meV/atom}$ and $35~\mbox{meV/atom}$, 
depending on the chirality of the tubes. The intertube distances are between $3.2~\mbox{\AA}$ 
and $3.4~\mbox{\AA}$ but independent of orientation for non C$_6$-axis tubes. 
\end{abstract} 


\maketitle 

\section{Introduction} 
Carbon is one of todays most exciting materials with properties originating from the 
structure and symmetry of its allotropes.~\cite{H.W.Kroto1985,RiichiroSaito1992,Iijima1993a,TeriWangOdom1998,K.S.Novoselov2004,RolandGillen2010} 
The properties of carbon nanotubes depend on their one dimensionality and the exact arrangement of 
the carbon atoms on their surface, called chirality.~\cite{RiichiroSaito1992,TeriWangOdom1998} 
Nanotubes are often found in bundles, ropes or fibers.~\cite{AndreasThess1996} These three dimensional 
structures form through the van der Waals interactions between the tube surfaces, which also influence 
the properties of the bundles.~\cite{MarcBockrath1997,S.Reich2002a} The orientation of tubes in a bundle 
can be compared to the stacking of graphene layers (e.g. bernal/AB stacked).~\cite{SusumuOkada2001,Chen2011} 
For nanotube bundles, however, we have to additionally consider the chirality of the tubes. The structural 
influence on the bundle properties is especially interesting for bundles of tubes that consist of only one 
chirality (monochiral bundles), as these have uniform properties needed for, e.g., electronic devices. 

The mechanical and electronic properties of bundles were already studied theoretically and 
experimentally.~\cite{Lu1997,Min-FengYu2000,Young-KyunKwon1998,PaulDelaney1999,AndreasThess1996,MinOuyang2001,AgnesSzabados2006} 
However, there is little knowledge about the influence of the structure (chirality) of the tubes on, e.g., 
the electronic structure of the bundle.~\cite{S.Reich2002a,SusumuOkada2001,HJLiu2007} Monochiral bundles, 
have not been experimentally produced yet, even so their production is expected in the near 
future.~\cite{ErikH.Haroz2010,JaredJ.Crochet2011,CarolinBlum2011} 

Nanotube bundles and their electronic structure were studied with density functional theory within the 
local density approximation (LDA), which allows to consider the structure of the tubes in the 
bundle.~\cite{S.Reich2002a,SusumuOkada2001,HJLiu2007} The local density approximation, however, fails 
in modeling the van der Waals interaction between tubes. Another approach is to use continuum 
approximations involving the Lennard-Jones potential. The continuum approximation, however, does not 
account for the specific configuration of the carbon atoms 
(structure).~\cite{LorinXBenedict1998,L.A.Girifalco2000,Cheng-HuaSun2006,AlexanderI.Zhbanov2010} Other 
approaches that consider the structure as well as the van der Waals interaction, do not reach the 
accuracy of density functional theory calculations.~\cite{RickF.Rajter2007,M.Seydou2011} The van der 
Waals functional developed by Dion~\emph{et al.} was shown to be able to model bundles of carbon 
nanotubes in density functional theory.~\cite{M.Dion2004,JesperKleis2008} At this level of theory 
studies considering the influence of nanotube chirality in bundles remain missing. 

In this paper we study the properties of bundles consisting of carbon nanotubes of the same chirality and 
handedness by a van der Waals density functional. We calculated the binding energy, intertube distance, 
and electronic structure in dependence of the orientation of the tubes in the bundle. We find a particularly 
strong dependence on the orientation of the tubes in the bundle if the tubes are achiral and the chirality 
of the tubes and the bundle share the trigonal symmetry, meaning that the tubes have a C$_6$-axis. Achiral 
tubes with C$_6$-axis are metallic, e.g., $\left(6,6\right)$ and $\left(12,0\right)$ tubes. 
Pseudogaps open at the Fermi level in dependence of the orientation of the tubes in the bundle. We find 
metallic behavior for the AA-stacked configuration ($0^{\circ}$) for C$_6$-axis armchair bundles, as well 
as for intermediate configurations which have glide reflection planes. The bundles of C$_6$-axis chiralities 
experience rotation barriers for the tubes that are induced by the configuration of the atoms on the 
tube surfaces. The barriers can be as high as $\Delta E=10.5~\mbox{meV}/\mbox{atom}$. Rotation barriers 
for tubes without the trigonal symmetry of the bundle are less than $\Delta E=0.3~\mbox{meV}/\mbox{atom}$. 
The properties of bundles made from non C$_6$-axis tubes have a much weaker dependence on the orientation 
of the tubes inside of the bundle, e.g. $\left(8,2\right)$ and $\left(14,0\right)$ tubes. 

This paper is organized as follows. We first describe the computational methods used for our study, 
Sec. II. We then discuss the van der Waals energies and intertube distances of bundles in orientational 
dependence in Sec. III. In Sec. IV we discuss the electronic band structure at the Fermi level for a 
$\left(6,6\right)$ bundle as a function of tube orientation. Section V summarizes this work and discusses 
the conclusions in a general context. 

\section{Computational methods} 
We used the \emph{ab-initio} package SIESTA to perform our density functional theory 
calculations.~\cite{PabloOrdejon1996,JoseSoler2002,G.Roman-Perez2009} To optimize the geometry of 
the isolated tubes we used the generalized gradient approximation parameterized by Perdew, Burke 
and Ernzerhof, which is good in modeling the interactions within the tube.~\cite{JohnP.Perdew1996} 
We calculated the bundle properties (binding strength, intertube distance and bundle electronic structure) 
with the van der Waals density functional parameterized by Dion~\emph{et al.}, which is especially good in 
modeling the van der Waals interaction between the tubes in the bundle.~\cite{M.Dion2004} Both calculations 
used norm conserving nonlocal pseudopotentials.~\cite{Troullier1991} We balanced the computational 
time and accuracy, by choosing localized pseudoatomic orbitals with a double-$\zeta$ (DZ) basis set 
to describe the valence electrons. We chose cutoff radii for the $s$ and $p$ orbital of the carbon 
atoms with $r_s=5.949~\mbox{Bohr}$ and $r_p=7.450~\mbox{Bohr}$. The mesh cutoff for the real-space 
integration corresponded to about $350~\mbox{Ry}$. We used between 10 and 14 $k$-points in the 
Monkhorst-Pack scheme~\cite{HendrikJ.Monkhorst1976} to calculate the total energies in the 
$k_z$-direction. The $z$-axis was chosen as the tube-axis. We performed all total energy calculations 
with only one $k$ point in the $x$ and $y$ direction. The $k$ point sampling for the band structure 
calculations of the bundled tubes were 20 $k$-points in $x$ and $y$ direction leading to a $k$ point 
sampling of 20x20x300 $k$-points for $\left(n,0\right)$ zigzag tubes, 20x20x300 for $\left(n,n\right)$ 
armchair tubes, and 20x20x200 for the chiral $\left(8,2\right)$ tube. In this paper we will only 
consider the band structure of the monochiral $\left(6,6\right)$ armchair tube bundle in detail. 

To obtain decent bundle structures we first optimized the lattice constant of isolated tubes by 
minimization of the total energy in dependence of the lattice constant. We then performed a geometry 
optimization for isolated carbon nanotubes within the generalized gradient approximation to derive 
an initial tube geometry for our bundled tube calculations. We used the conjugate gradient method 
to optimize the atomic coordinates to a maximal force tolerance of $0.04~\mbox{eV/\AA}$, which is 
adequate for our accuracy. The optimized isolated tube coordinates were then used to perform the 
van der Waals density functional calculations. 

\begin{figure}
\includegraphics[scale=0.15]{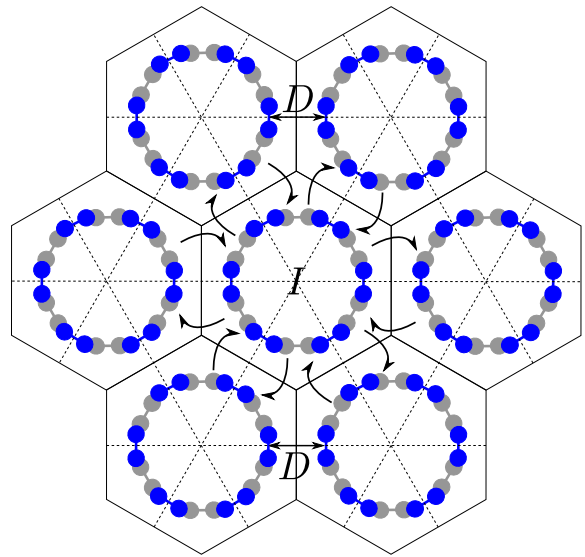} %
\caption{\label{hexagonaluc}Ball and stick sketch of a bundle of seven $\left(6,6\right)$-tubes in their high-symmetry orientation ($0^{\circ}$). The hexagon in the middle (inner tube $I$) represents the unit cell for our bundle calculations. Lines connecting the middle points of the hexagons were added for clarity. The top carbon atoms are highlighted in blue, bottom carbon atoms are gray. The symmetry for seven tubes is trigonal. The inner tube $I$ interacts with six neighbors with intertube distance $D$.}
\end{figure}

For the van der Waals density functional calculations we placed a tube in a hexagonal unit cell with periodic 
boundaries, see Fig.~\ref{hexagonaluc} for the example of a bundle of $\left(6,6\right)$-tubes. We calculate 
a bundle by an infinite number of tubes (bulk-bundle), which serves as a model of an inner tube $I$ in a real 
carbon nanotube bundle.

\begin{figure}
\includegraphics[scale=0.50]{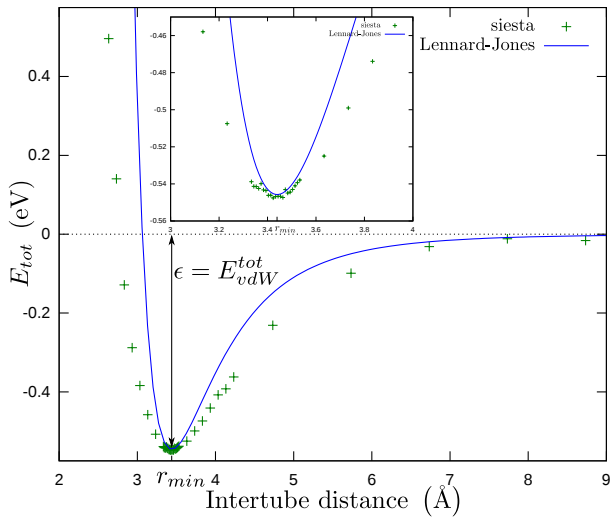}
\caption{\label{66EvsD}Total energy in dependence of the intertube distance calculated for a bundle of $\left(6,6\right)$ tubes. The energy is normalized to the energy of an isolated tube, which corresponds to infinitely distant neigboring tubes. The data points were obtained by SIESTA. A Lennard-Jones potential is plotted with $\epsilon$ and $r_{min}$ parameters fitted to our data.} 
\end{figure}

The total energy of the bulk bundle was minimized in dependence of the intertube distance to obtain the optimal 
distance between the tubes of the bundle as presented in Fig.~\ref{66EvsD}. A Lennard-Jones potential is plotted 
with $\epsilon$ and $r_{min}$ parameters fitted to our data. The Lennard-Jones potential shows a smaller width 
in the attractive region compared to the calculated data points. We did not perform a structure relaxation with 
the van der Waals density functional as this leads to energetically lowest states, which does not allow to 
consider energetically unstable orientations of the tubes of the bundle. Therefore, stress remains in the bundle 
systems depending on the chirality and to a smaller extent on the orientation of the tubes. The maximal forces 
in the systems are $F_{max}\approx0.3~\mbox{eV/\AA}$ for armchair bundles, $F_{max}\approx1~\mbox{eV/\AA}$ for 
zigzag bundles, and $F_{max}\approx3~\mbox{eV/\AA}$ for the $\left(8,2\right)$ and $\left(12,6\right)$ bundle. 
Previous studies of tube-tube interaction found the distortion caused by the bundling to have little or no 
effect on the properties for tubes with diameters below $15~\mbox{\AA}$.~\cite{RodneyS.Ruoff1993,S.Reich2002a} 

The total van der Waals energy is defined as the difference between minimal total energy of 
the bundle and total energy of the isolated tube (both calculated with the functional of 
Dion~\emph{et al.}~\cite{M.Dion2004}), which corresponds to infinite intertube distance. Dividing 
through the number of atoms of the unit cell yields the van der Waals energy per atom. We derive 
the intertube distance from the optimized $xy$-unit cell length by subtraction of the diameter 
of the tube. The band structure was plotted for certain high symmetry directions. We used the whole 
$k$-sampled zone to derive the density of states at an electronic temperature of $20~\mbox{K}$. 

\section{Van der Waals energies and intertube distances} 
We start this section with a discussion of the symmetry properties of nanotube bundles. This 
discussion is followed by a study of the influence of the orientation of tubes in the bundle 
on the van der Waals energies between the tubes of the bundle, as well as the energetically 
optimized intertube distances. At the end of the section we draw a conclusion from our results 
for the properties of mixed chirality bundles and monochiral bundles. 

Bundles were experimentally observed in a triangular lattice that contained up to hundreds 
of tubes.~\cite{AndreasThess1996,J.M.Cowley1997} In our calculations we model the triangular 
lattice of the bundle by a hexagonal unit cell, compare Fig.~\ref{hexagonaluc}. A symmetry 
breaking occurs, if the individual tube and the bundle do not share all symmetry operations. 
The bundle structure has in general a D$_{6h}$ symmetry, this means it has symmetry axes 
(2 C$_6$, 2 C$_3$, C$_2$, 3 C'$_2$, 3 C''$_2$), mirror planes (3 $\sigma_v$, 3 $\sigma_d$, 
$\sigma_h$), one inversion center ($i$), rotation-reflection axes (2 S$_3$, 2 S$_6$) and one 
identity element ($E$). The tubes have symmetry operations depending on their chirality. The 
level of symmetry breaking depends on the number of shared symmetry elements between the bundle 
structure and the chirality of the individual tube. A tube in the bundle has to have the same 
atom configuration every $60\,^{\circ}$ on the circumference of the tube to share symmetry operations 
with the bundle. If the chirality and bundle structure share the symmetry, we say that the chirality 
has a C$_6$-axis and symmetry breaking is lifted in high symmetry configurations (e.g. $0^{\circ}$ 
in Fig.~\ref{hexagonaluc}). The high symmetry configuration has all symmetry operations of the 
D$_{6h}$ symmetry, especially mirror planes. 

We study the orientational influence on the properties of the bundles by simultaneous rotation 
of all tubes of the bundle in steps of $1\,^{\circ}$ or $5\,^{\circ}$ starting from 
$0\,^{\circ}$ to $60\,^{\circ}$. We consider the $\left(6,6\right)$, $\left(12,12\right)$, 
$\left(9,0\right)$, $\left(12,0\right)$ and $\left(12,6\right)$ C$_6$-axis chiralities. The 
$\left(12,0\right)$-tube structure, for example, has the same atomic configuration every 
$15\,^{\circ}$ around the circumference (considering screw operations), as the full circumference 
contains $360\,^{\circ}$ and there are 24 atom positions on the circumference of the tube. This 
also means that the $\left(12,0\right)$-tube has the same atomic configuration at the angles 
$0\,^{\circ}$/$360\,^{\circ}$, $60\,^{\circ}$, $120\,^{\circ}$, $180\,^{\circ}$, $240\,^{\circ}$ 
and $300\,^{\circ}$, therefore it has a C$_6$-axis. To study the effect of intertube orientation 
on non C$_6$-axis chiralities we further study the chiral $\left(8,2\right)$-bundle, the two 
zigzag bundles $\left(8,0\right)$ and $\left(14,0\right)$ and the two armchair bundles 
$\left(5,5\right)$ and $\left(10,10\right)$, which have nearly no common symmetry elements with 
the bundle structure. 

\begin{table}
\caption{\label{rotationbundle}Comparison of the intertube distance $D$, van der Waals energy per atom $|E_{vdW}^{atom}|$, and their variations $\Delta \left(D, E_{vdW}^{atom}\right)$ for various chiralities $\left(n,m\right)$, chiral angles $\theta$ and tube diameters $d$. The table presents the dependence of the fundamental properties on the orientation (rotation) of the tubes in the bundle with minimal and maximal values. Errors for intertube distances are $\pm 0.02~\mbox{\AA}$ and for binding energies the errors are $\pm 0.2~\mbox{meV/atom}$.} 

\begin{tabular}{ccccccc}
$\left(n,m\right)$ & $\theta$~$\left(^{\circ}\right)$ & $d$~$\left(\mbox{\AA}\right)$ & $D$~$\left(\mbox{\AA}\right)$ & $\Delta D$ & $|E_{vdW}^{atom}|$~$\left(\mbox{meV}\right)$ & $\Delta E_{vdW}^{atom}$ \\
\hline
$\left(8,0\right)$   &  0   & 6.41  & 3.29-3.32 & 0.03 & 31.1-31.3 &  0.2 \\
$\left(5,5\right)$   & 30   & 6.92  & 3.33-3.35 & 0.02 & 29.5-29.6 &  0.1 \\
$\left(9,0\right)$   &  0   & 7.19  & 3.17-3.35 & 0.18 & 29.5-34.7 &  5.2 \\
$\left(8,2\right)$   & 10.9 & 7.31  & 3.29-3.32 & 0.03 & 29.3-29.6 &  0.3 \\
$\left(6,6\right)$   & 30   & 8.27  & 3.15-3.44 & 0.29 & 22.7-33.2 & 10.5 \\
$\left(12,0\right)$  &  0   & 9.53  & 3.25-3.40 & 0.15 & 23.9-26.9 &  3.0 \\
$\left(14,0\right)$  &  0   & 11.10 & 3.31-3.33 & 0.02 & 23.8-24.0 &  0.2 \\
$\left(12,6\right)$  & 19.1 & 12.56 & 3.26-3.30 & 0.04 & 21.6-22.5 &  0.9 \\
$\left(10,10\right)$ & 30   & 13.70 & 3.27-3.31 & 0.04 & 22.2-22.5 &  0.3 \\
$\left(12,12\right)$ & 30   & 16.42 & 3.22-3.37 & 0.15 & 18.8-21.5 &  2.7 \\
\hline
\end{tabular} 

\end{table} 

We present the minimal values, maximal values and variation in dependence of the 
orientation of the tubes (rotation angle) for the intertube distance D and van 
der Waals energy per atom $|E_{vdW}^{atom}|$ for various chiralities and tube diameters, 
see Tab.~\ref{rotationbundle}. We only see small variations for the values of non 
C$_6$-axis chiralities, e.g. for the bundle of $\left(8,2\right)$-tubes. The intertube 
distance of the $\left(8,2\right)$-bundle varies by $\Delta D\approx0.03~\mbox{\AA}$ and 
$E_{vdW}^{atom}$ varies by $\approx0.3~\mbox{meV}$, meaning by $1\%$ or less. Our results 
correspond well to results of the $\left(10,10\right)$-bundle, with our maximal difference 
between lowest and highest van der Waals energy per atom $\Delta E_{vdW}=0.3~\mbox{meV}$ 
and the activation barrier for rotations of $\Delta E_{vdW}=0.15~\mbox{meV}$ reported 
previously.~\cite{Young-KyunKwon1998} The activation barrier results from the reduction 
in symmetry from D$_{2h}$ to C$_{2h}$ due to rotation.~\cite{PaulDelaney1999} The 
properties of bundles made from non C$_6$-axis chiralities show hardly any dependence on 
orientation. 

Comparing the intertube distance for the $\left(8,0\right)$ and the $\left(14,0\right)$ tube, 
we find a variation of less than $1\mbox{-}2\%$, whereas the tube diameter increases by 
$\approx73~\%$. The van der Waals energy per atom decreases about $23~\%$ for the same 
diameter comparison. The tube diameter has a strong influence on the binding strength, 
but does not influence the intertube distance. 

For the C$_6$-axis chiralities, we receive a radically different result for the 
dependence of the properties of the bundles on the rotation angle. We observe a 
variation of the bundle properties in dependence of the orientation for all C$_6$-axis 
chiralities, compare Tab.~\ref{rotationbundle}. Starting with the zigzag chirality 
$\left(9,0\right)$; The intertube distance varies by $\Delta D\approx0.18~\mbox{\AA}$, 
corresponding to about $6~\%$ variation, which is six times higher than the variation 
observed for non C$_6$-axis chiralities and at least three times higher than the variation 
associated with tube diameter. Furthermore $\Delta E_{vdW}^{atom}\approx5.2~\mbox{meV}$, 
corresponding to about $15~\%$ variation, which is 15 times higher than the variation 
observed for non C$_6$-axis chiralities and only slightly smaller than the variation 
accounted to the tube diameter. For the $\left(12,0\right)$ chirality the orientational 
dependence on the properties weakens, with $\Delta D\approx0.15~\mbox{\AA}$ $\left(\approx4\%\right)$ 
and $\Delta E_{vdW}^{atom}\approx3~\mbox{meV}$ $\left(\approx11\%\right)$. The rotation 
energy barrier $\Delta E_{vdW}^{atom}\approx3.0~\mbox{meV}$ corresponds well to the 
barrier of $3.0~\mbox{meV}$ previously reported by LDA calculations.~\cite{SusumuOkada2001}
The influence of the orientation on the bundle properties decreases with increasing tube 
diameter limiting the occurrence of the special properties of the C$_6$-axis bundles to 
small diameter nanotubes. We suppose that the increase in interaction area lowers the 
influence of the local symmetry. The interaction area increases through the reduced 
curvature of larger-diameter tubes. 

The $\left(6,6\right)$ armchair bundle shows the strongest orientation dependence with 
$\Delta D\approx0.29~\mbox{\AA}$ $\left(\approx9\%\right)$. The rotation energy barrier 
$\Delta E_{vdW}^{atom}\approx10.5~\mbox{meV}$ $\left(\approx32\%\right)$ is larger than 
$\approx5.0~\mbox{meV}$ previously reported by LDA calculations.~\cite{SusumuOkada2001} 
Interestingly, the variations of the bundle parameters in dependence of the orientation 
are smaller for the zigzag bundles than for the armchair bundles of comparable tube 
diameter. 

For C$_6$-axis bundles of chiral tubes (e.g. $\left(12,6\right)$) we see a dependence on 
the van der Waals energy ($\Delta E_{vdW}^{atom}\approx0.9~\mbox{meV}$ 
$\left(\approx4\%\right)$) and the intertube distance ($\Delta D\approx0.04~\mbox{\AA}$ 
$\left(\approx1\%\right)$), however, the dependence is low. 
In chiral C$_6$-axis bundles alignment of neighboring tubes is impossible due to the 
handedness of chiral tubes. Neighboring tubes with same handedness have opposing surfaces 
with different handedness leading to mismatched surface structures of neighboring surface 
atom layers. It is not possible to interchange handedness of neighboring tubes to generate 
alignment for all neighbors as neighbors with same handedness remain caused through the 
trigonal structure of the bundle. 

\begin{figure}
\includegraphics[scale=0.50]{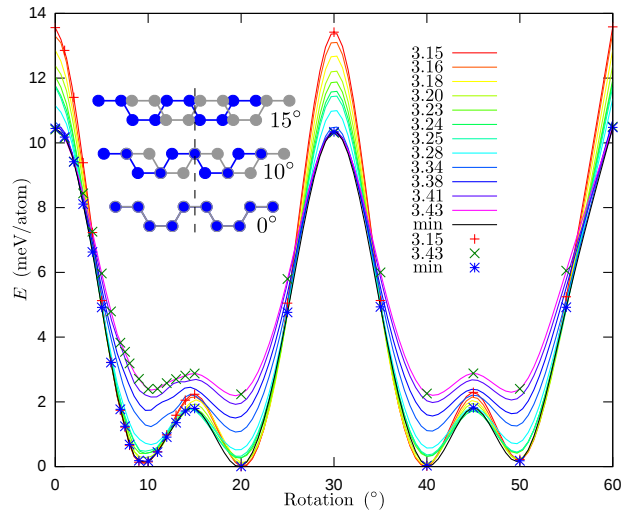} %
\caption{\label{Erotminmax}Total energy for various intertube distances as a function of rotation angle for a bundle of $\left(6,6\right)$-tubes. The total energy depends on the intertube distance as well as the orientation of the tubes. The inset shows a side view sketch of the atomic configuration for the three most interesting rotation angles. The front tube has blue atoms, the tube in the back is in gray.} 
\end{figure}

The intertube distance of achiral C$_6$-axis nanotubes depends on their orientation and 
varies by as much as $10\%$. The binding strength of C$_6$-axis bundles is influenced 
through the variation of the intertube distance but at the same time by the orientation; 
see Fig.~\ref{Erotminmax}. The influence of the intertube distance on the total energy has 
a strong impact for $0^{\circ}$ (AA-stacked) and around $10^{\circ}$ (AB-stacked), but is 
weaker around $4^{\circ}$. The orientational influence on the binding strength of C$_6$-axis 
chiralities with small diameters ($<20~\mbox{\AA}$) is of the same magnitude as the influence 
of the tube diameter on the binding strength in the bundle. 

\begin{figure}
\includegraphics[scale=0.50]{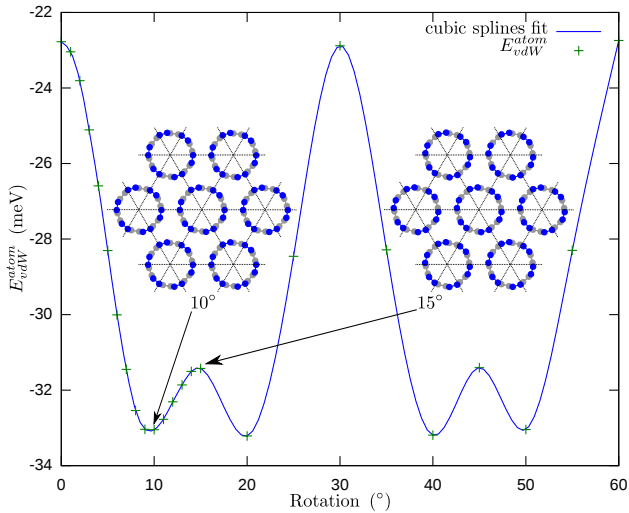} %
\caption{\label{FE66rot}Van der Waals energy per atom $E_{vdW}^{atom}$ as a function of rotation angle for a bundle of $\left(6,6\right)$-tubes. The tubes of the bundle are rotated starting from the high symmetry position $\left(0\,^{\circ}\mbox{, see Fig.~\ref{hexagonaluc}}\right)$. The lines are cubic spline fits and show symmetry breaking behavior. Global extrema occur every $10\,^{\circ}$. Minimum values occur at $10\,^{\circ}+30\,^{\circ}\cdot n$ and $20\,^{\circ}+30\,^{\circ}\cdot n$ with integer $n$. Global maxima occur every $30\,^{\circ}$ starting from $0\,^{\circ}$ and local maxima occur every $30\,^{\circ}$ starting from $15\,^{\circ}$. The insets show the two exemplary configurations of $10\,^{\circ}$ and $15\,^{\circ}$ rotation of the tubes in the bundle. The blue spheres mark the position of the top carbon atoms, the gray spheres mark the position of the bottom carbon atoms.} 
\end{figure}

We use the example of the $\left(6,6\right)$-bundle to discuss the physical properties of 
a C$_6$-axis bundle. In this bundle the properties depend most strongly on rotational 
orientation. The van der Waals energy per atom of the $\left(6,6\right)$-bundle 
as a function of rotation angle shows symmetry breaking behavior, see Fig.~\ref{FE66rot}. 
Symmetry is initially D$_{6h}$ ($0\,^{\circ}$) and reduces to C$_{6h}$ (loss of mirror planes) 
due to rotation. The high symmetry configuration of $0\,^{\circ}$, corresponding to AA stacking 
in graphite~\cite{SusumuOkada2001}, is not energetically stable, see Fig.~\ref{Erotminmax}. 
It has one of the highest intertube distances of $3.43~\mbox{\AA}$. The magnitude of the 
van der Waals energy increases up to a maximum at $\approx10\,^{\circ}$, corresponding to 
AB stacking in graphite~\cite{SusumuOkada2001}, see Fig.~\ref{FE66rot}. This is in contrast 
to previous LDA findings, where a configuration about $2.5^{\circ}$ off the AB stacking lead 
to a maximum at about $7.5\,^{\circ}$.~\cite{SusumuOkada2001} The configuration at 
$10\,^{\circ}$ has one of the smallest intertube distances observed in our calculations with 
$3.15~\mbox{\AA}$. At $15\,^{\circ}$ the rotation leads to an interesting configuration that 
has glide reflection planes, with a local binding energy minimum and a moderate intertube 
distance of $3.25~\mbox{\AA}$. Further rotations only reproduce the behavior that is contained 
in the first $15\,^{\circ}$ rotation, see Fig.~\ref{FE66rot}. 

The error in our calculation can be estimated by comparing two identical configurations, e.g. 
$10\,^{\circ}$, $20\,^{\circ}$, $40\,^{\circ}$, and $50\,^{\circ}$ in Fig.~\ref{FE66rot}. We 
obtain an error of $\pm 0.2~\mbox{meV/atom}$ for the total energy. An error of 
$\pm 0.02~\mbox{\AA}$ for the intertube distance can be estimated from the data in 
Fig.~\ref{Erotminmax}. 

The smaller activation barrier for rotations between $15\,^{\circ}$ and $20\,^{\circ}$ has 
a value of $\Delta E=1.8~\mbox{meV}/\mbox{atom}$ and the second, larger activation barrier 
for rotations between all orientations (e.g. between $0\,^{\circ}$ and $10\,^{\circ}$) is 
$\Delta E=10.5~\mbox{meV}/\mbox{atom}$. Rotations of a solid made from $C_{60}$ fullerenes 
were experimentally and theoretically found to be hindered below $T\approx 260~\mbox{K}$, 
implicating that a hinderance of rotations for nanotubes in a bundle might also be 
possible.~\cite{C.S.Yannoni1991,R.Tycko1991,JianPingLu1992,Young-KyunKwon1998} 

At the end of this section we want to discuss our findings and the implications for mixed chirality 
carbon nanotube bundles. Most carbon nanotube bundles contain nanotubes of various chiralities 
$\left(n,m\right)$. The binding strength and intertube distance depend in general on the orientation 
of the tubes in the bundle. But the orientational dependence of the aforementioned properties is 
suppressed by symmetry breaking induced by mismatch of the bundle and tube symmetry. We can conclude 
from our calculations, that the properties of bundles of mixed chiralities have a negligible dependence 
on tube orientation. This results from the non C$_6$-axis chirality observations, which show only a 
weak dependence on the orientation for the bundle properties (about or less than $1\%$ variation 
in orientation dependence). The tube diameter (curvature) otherwise has a strong influence on the 
binding strength, but does not influence the intertube distance. Recently Crochet~\emph{et al.} 
succeeded in producing nearly mono chirality single walled carbon nanotube bundles, highlighting 
the possibility to produce monochiral samples in the near future which would allow to experimentally 
test the theoretically predicted properties.~\cite{ErikH.Haroz2010,JaredJ.Crochet2011,CarolinBlum2011} 
Only the monochiral C$_6$-axis tubes preserve the symmetry in the higher symmetry configurations. 
Chiralities with C$_6$-axis can exhibit special properties, e.g. binding energy and intertube distance 
are influenced in orientational dependence in contrast to the tube diameter which only influences the 
binding energy. 

\section{Electronic band structure at the Fermi level} 
In this section we study the band structure, density of states, and pseudogap and 
their variation in bundles of metallic tubes belonging to the C$_6$-axis chirality 
at the example of $\left(6,6\right)$-tubes. At the end of the section we compare 
our results to other theoretical and experimental studies and discuss the influence 
of rotation barriers on the electronic properties of bundles. 

\begin{figure}
\includegraphics[scale=0.50]{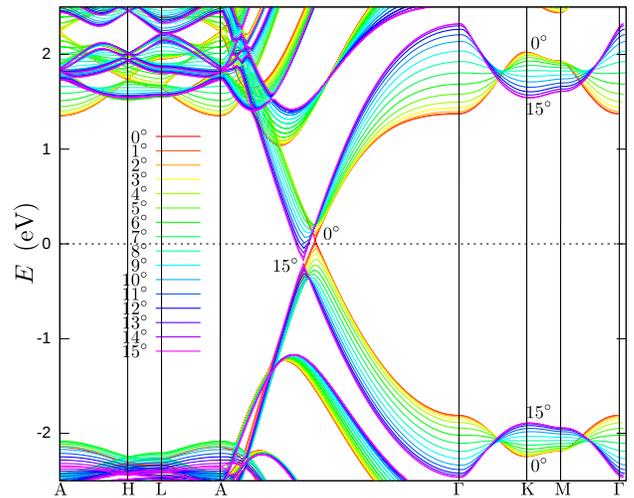}
\caption{\label{66rotbandsfull}Band structure along high symmetry directions of a bundle of $\left(6,6\right)$-tubes rotated starting from its high symmetry configuration $\left(0\,^{\circ}\right)$ in rainbow colors (color online). The band structure diagrams are normalized to the Fermi levels.}
\end{figure} 

\begin{figure}
\includegraphics[scale=0.50]{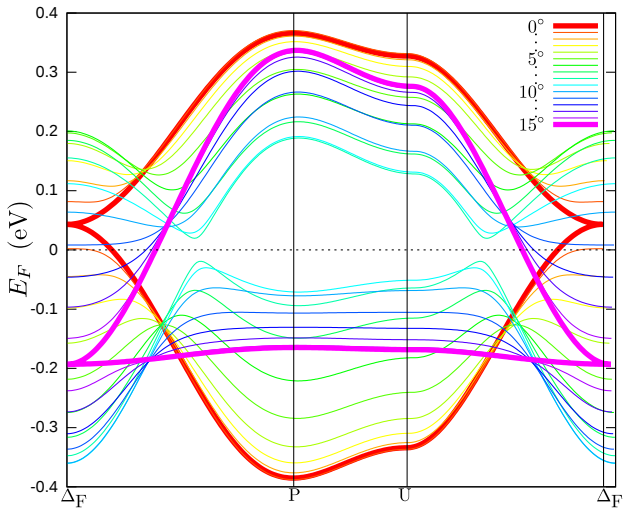}
\caption{\label{deltaF}Band structure along high symmetry directions of a bundle of $\left(6,6\right)$-tubes starting from the position of the valence and conduction band crossing in rainbow colors (color online). The band structure diagrams are normalized to the Fermi levels. The valence band flattens by change in orientation while the influence on the conduction band is smaller and it mostly keeps its shape. The bands for the orientations with metallic behavior (without pseudogap) are highlighted.} 
\end{figure} 

Band structure diagrams of a bundle of $\left(6,6\right)$-tubes for orientations 
of $0^\circ$ to $15^\circ$ show that the bands change their slope with orientation, 
see Fig.~\ref{66rotbandsfull}. The valence band maximum and conduction band minimum 
shift in energy- and $k$-direction through the change in orientation. This is 
accompanied by a shift of the Fermi level, as can be seen between A and $\Gamma$ 
(all Fermi levels were normalized to 0 eV). The Fermi levels shift through the 
changed electron density of each orientation, which result from a volume change 
induced by the orientational specific intertube distances. For the high symmetry 
configuration ($0^{\circ}$, D$_{6h}$) the valence band and conduction band cross at about 
$43~\mbox{meV}$ above the Fermi level, which corresponds well to the value of about 
$70~\mbox{meV}$ calculated within LDA calculations; the Fermi level is shifted 
compared to the Fermi level of the isolated tube.~\cite{S.Reich2002a} The $k_z$ 
value for the position of the valence band maximum and conduction band minimum is 
between $k_z=0.60\cdot\frac{\pi}{a}$ for $0\,^{\circ}$ and 
$k_z=0.66\cdot\frac{\pi}{a}$ for $15\,^{\circ}$, where $a=2.491~\mbox{\AA}$ is 
the lattice constant along the tube. Our $k_z$-values correspond well to the value 
in isolated tubes of $k_z=2/3\cdot\frac{\pi}{a}$. There is no band splitting in 
the orientations at $0\,^{\circ}$ and $15\,^{\circ}$. The different parity of the 
bands allows the crossing of the bands. The band splits by rotating the tubes of the 
bundle as little as $1\,^{\circ}$ out of the high symmetry position, which opens up 
a band gap of $E_g=75~\mbox{meV}$ along the A$\Gamma$-direction. The mirror symmetry 
for rotated configurations is broken, which leads to anticrossing. The band gap 
increases until $\approx7\,^{\circ}$ and then becomes smaller again until it vanishes 
at $15\,^{\circ}$. The valence band flattens in dependence of the orientation for the 
high symmetry directions starting from the valence and conduction band crossing, see 
Fig.~\ref{deltaF}. 

\begin{figure*}
\includegraphics[scale=0.50]{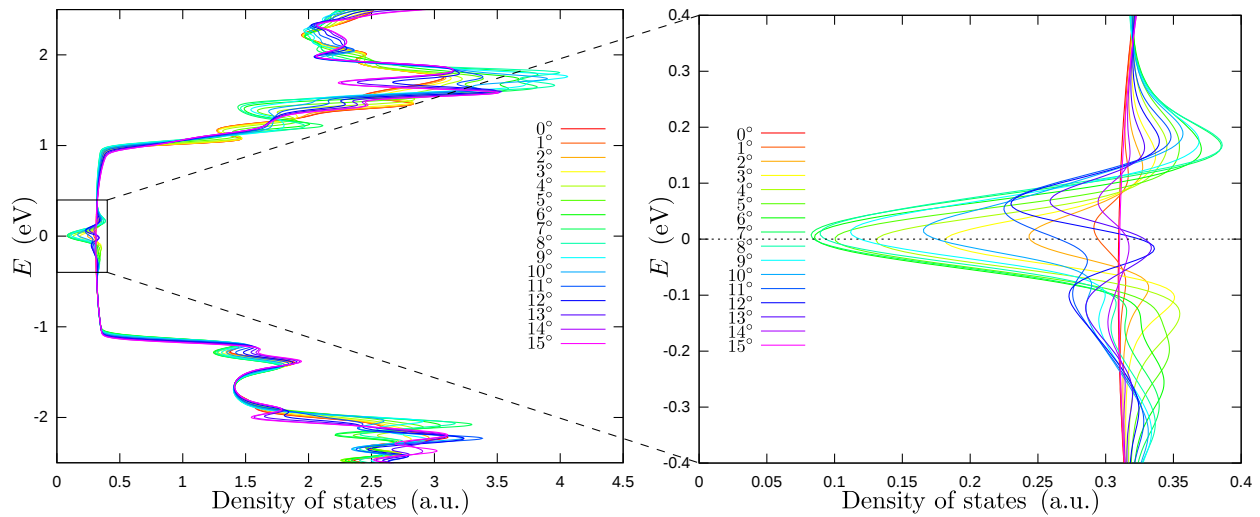} 
\caption{\label{66rotdos}Density of states of a bundle of $\left(6,6\right)$-tubes rotated starting from its high symmetry position $\left(0\,^{\circ}\right)$ in rainbow colors (color online). The density of states diagrams are normalized to the Fermi levels. The right panel shows the shift of the pseudogap minimum with respect to the Fermi level ($0~\mbox{eV}$), which can also be observed in Fig.~\ref{66rotbandsfull}.}
\end{figure*} 

The high symmetry configuration ($0^{\circ}$) shows metallic behavior while 
an increase of rotation angle of the tubes in the bundle increases a pseudogap, 
which is maximal at $\approx7^\circ$, see density of states in Fig.~\ref{66rotdos}. 
The pseudogap diminishes with further rotation until it closes at $15\,^{\circ}$, 
which can be accounted for by the symmetry of the configuration. The $15\,^{\circ}$ 
configuration looses mirror planes, but has glide reflection planes, which keeps 
the configuration metallic, see Figs.~\ref{deltaF} and \ref{66rotdos}. Furthermore, 
the density of states minima shift around the Fermi level from $0~\mbox{meV}$ at 
the global minimum to $50~\mbox{meV}$ in dependence of the orientation, see right 
panel in Fig.~\ref{66rotdos}. The shift of the density of states maxima leads to 
increased density of states at the Fermi level for orientations between 
$12\,^{\circ}$ and $14\,^{\circ}$, compared to the metal like behavior of the 
$0\,^{\circ}$ orientation. This can partly be understood by comparison to the 
behavior observed by the band structure, see Fig.~\ref{66rotbandsfull}. 

The orientation has a smaller influence on the valence side than on the conduction 
side of the band structure, see Fig.~\ref{66rotdos}. The first peak of the conduction 
side shifts between 1.07 eV and 1.22 eV through rotation; for certain rotation 
angles (e.g. $15^\circ$) it becomes a shoulder. The valence band shows only 
small variation for the first two peaks. 

We conclude that the most exciting properties of C$_6$-axis bundles can be found in the 
band structure and density of states. Achiral C$_6$-axis bundles can be metallic. Certain 
orientations possess higher density of states at the Fermi level than the metallic 
configurations. The minima of the density of states meanwhile shift around the Fermi 
level in dependence of the orientation. The band structure shows a dive of the conduction 
band minimum below the Fermi level along the high symmetry direction accompanied by a 
pseudogap opening which closes again for higher symmetry configurations that have mirror 
planes or glide reflection planes. 

We now want to compare our results to other theoretical studies. The density of states of 
the high symmetry and rotated $\left(10,10\right)$ monochiral nanotube bundles were shown 
to have very similar behavior, which agrees with our results.~\cite{PaulDelaney1999,Young-KyunKwon1998} 
For C$_6$-axis bundles, however, we find a strong coupling between the electronic states near 
the Fermi level and the rotational motion, which was suggested to lead to superconducting 
behavior previously.~\cite{Young-KyunKwon1998} In contrast to previous studies we find 
metallic behavior for achiral C$_6$-axis bundles (e.g. $\left(6,6\right)$, $\left(12,12\right)$) 
for multiple orientations.~\cite{SusumuOkada2001} We were able to show, that no energy gap, 
but a pseudogap opens in dependence of orientation of the tubes in the bundle in contrast 
to previous studies, which showed an energy gap in the density of states for AB-stacked 
$\left(6,6\right)$ bundles as well as for the $8^{\circ}$ orientation.~\cite{SusumuOkada2001} 
Furthermore our results show a shift of the density of state extrema around the Fermi level, 
corresponding to previously reported results.~\cite{PaulDelaney1999} We also find, however, 
an increased density of states at the Fermi level compared to the metallic configurations 
for certain orientations. We find good agreement in comparison to experimentally derived density 
of states with the general trend of pseudogaps opening due to the bundling.~\cite{MinOuyang2001} 

At the end of the section we want to discuss the influence of rotation barriers on the electronic 
properties of monochiral achiral C$_6$-axis bundles. At low temperatures, certain configurations 
of monochiral C$_6$-axis chirality bundles can be stable, e.g. the orientation at $\approx10^\circ$ 
for the $\left(6,6\right)$-bundle. At room temperature thermal energy is likely to lead to rotations 
and vibrations of the tubes in the bundle. A complex electronic behavior is expected, as rotations 
(orientational changes) and absorption and emission occur time averaged over the experiment. Especially 
the slope flipping behavior and shifts of the bands lead to a further broadening of the width of the 
bands, which further broadens the density of states and therefore also the optical-absorption bands for 
the C$_6$-axis chiralities. The high symmetry configuration with metallic behavior is the energetically 
most unstable configuration and therefore semi-metallic behavior has to be expected for bundles of 
metallic tubes. The $15^{\circ}$ configuration is a local total energy maximum and therefore also 
expected to be an unstable configuration, even so it is much lower in energy than the high symmetry 
configuration. For larger diameter tubes the rotation barriers flatten, allowing metallic behavior 
in C$_6$-axis bundles of large diameter armchair tubes. 

\section{Conclusions}
In summary we presented van der Waals energies and intertube distances of various chiralities of 
carbon nanotube bundles in dependence of the orientation of the tubes inside of the bundle. 
Furthermore the electronic structure of the monochiral $\left(6,6\right)$-bundle was studied in 
dependence of the tube orientation. For tubes with a C$_6$ axis the orientation of the tubes in 
the bundle becomes a new degree of freedom to adjust the bundle properties. The intertube distance 
as well as the binding strength vary in dependence of the tube orientation. This dependence 
decreases with increasing tube diameter. Therefore this effect is mainly important for bundles 
composed of small diameter tubes with high curvature. By variation of the orientation of the tubes 
a pseudogap opens and increases until it vanishes at the next configuration which preserves the 
symmetry. C$_6$-axis bundles of armchair tubes have metallic configurations. Certain configurations 
show a higher density of states at the Fermi level than the metallic configurations. As recent 
progress suggests monochiral C$_6$-axis bundles will soon be experimentally available, which will 
give access to study the newly arisen bundle properties. 

\begin{acknowledgments}
We acknowledge F. Hennrich, S. Heeg, A. Setaro, C. Lehmann and B. Hatting for useful discussions. 
This work was supported by ERC under grant no. 210642. 
\end{acknowledgments}

\end{document}